# Inelastic Electron Tunneling Spectroscopy at High-Temperatures

*Prosper Ngabonziza[1], Yi Wang[1], Peter A. van Aken[1],*

*Joachim Maier[1], and Jochen Mannhart[1,*]*

[1]Max Planck Institute for Solid State Research,

Heisenbergstraße 1, 70569 Stuttgart, Germany

## Abstract

Ion conducting materials are critical components of batteries, fuel cells, and devices such as memristive switches. Analytical tools are therefore sought that allow the behavior of ions in solids to be monitored and analyzed with high spatial resolution and in real time. In principle, inelastic tunneling spectroscopy offers these capabilities. However, as its spectral resolution is limited by thermal softening of the Fermi–Dirac distribution, tunneling spectroscopy is usually constrained to cryogenic temperatures. This constraint would seem to render tunneling spectroscopy useless for studying ions in motion. We report here the first inelastic tunneling spectroscopy studies above room temperature. For these measurements, we have developed high-temperature-stable tunnel junctions that incorporate within the tunnel barrier ultrathin layers for efficient proton conduction. By analyzing the vibrational modes of O–H bonds in $BaZrO_3$-based heterostructures, we demonstrate the detection of protons with a spectral resolution of 20 meV at 400 K (FWHM). Overturning the hitherto existing prediction for the spectral resolution limit of 186 meV (5.4 $k_B T$ at 400 K), this resolution enables high-temperature tunneling spectroscopy of ion conductors. With these advances, inelastic tunneling spectroscopy constitutes a novel, valuable analytical tool for solid-state ionics.

---

[*] Corresponding author: office-mannhart@fkf.mpg.de

# Introduction

Inelastic tunneling spectroscopy (IETS) is an analytical technique to monitor and analyze accurately the diffusion processes of particles in solids. IETS is highly sensitive and offers temporal resolution plus species specificity. It is used with outstanding success to identify and explore ions and other particles, defects, and elementary excitations in tunnel junctions.[1,2,3,4,5,6,7,8,9] By its very nature, IETS is capable of detecting particles in ultra-thin barriers or even interfaces. Other spectroscopic techniques often used to characterize ion conductors such as infrared (IR) spectroscopy, Raman spectroscopy, inelastic neutron scattering (INS), and Rutherford backscattering lack this efficiency.[10,11,12] Moreover, IETS is sensitive to both IR and Raman-active transitions. Even the optically forbidden regions that are inaccessible to IR and Raman spectroscopy are open to IETS,[2,8] making it a promising technique for solid-state ionics. If available, it would open the door to monitoring the simultaneous propagation of one or several particle species in solids separately and in real time. IETS is widely used with great success as a spectroscopy technique for scanning tunneling microscopy.[5,9,13,14] This variant of IETS is called inelastic scanning tunneling spectroscopy. We focus here on IETS performed by using planar tunnel junctions, which offer multichannel, on-chip integration.

Despite all these attractive properties, IETS has been disregarded as a technique in solid-state ionics to analyze particle motion because the temperatures necessary to study diffusion processes are high, usually well exceeding 300 K. These temperatures were understood to be out of range for IETS. As early as 1968, J. Lambe and R. C. Jaklevic derived that, due to the thermal softening of the tunnel contacts' Fermi–Dirac distributions, the limit of the spectral resolution of IETS worsens linearly with temperature $T$, the best possible resolution being 5.4 $k_B T$, where $k_B$ is the Boltzmann's constant.[2] Measurements of the resolution limit of Al–Al$_2$O$_3$–Pb junctions between 4.2 and 148 K are consistent with this limit.[15] Consequently, IETS has been widely considered to be useless above 150 K.[1,2,3,16] Room-temperature electron tunnelling currents have been employed successfully to identify single amino acids generated by cleaving proteins.[17] In that work, conductance versus time profiles of the molecules moving through tunnel junctions, were used to identify the molecules, thereby averting the need for spectroscopical resolution. We are aware of only two publications that include room-temperature IETS spectra;[18,19] indeed room temperature appears to be the highest temperature at which IETS has ever been tried. The temperature gap between IETS and solid-state ionics has also blocked the use of IETS as an analytical tool for nanoionic studies.

However, we reveal here that the limit of 5.4 $k_B T$ is not valid at temperatures $T$ above 50 K. At these temperatures, the achievable spectral resolution of IETS can be a factor of 9 better, with no limit found yet. We have obtained a minimal spectral width (FWHM) of 19 meV at 300 K and 20.1 meV at 400 K. Consequently, useful IETS can be performed at room temperature and even above,[19] as we have verified experimentally by our IETS-results relating to protons diffusing along interfaces in oxide heterostructures up to 400 K. The capability of high-temperature operation proposes IETS as a tool for nanoionics and nanoelectronics.

The operating principle of IETS is illustrated in **Figure 1**a. Electrons with an initial energy $E_i$ are found to tunnel elastically or inelastically across a barrier. In an inelastic tunneling process, a tunneling electron excites a quasiparticle such as a phonon, an ion, an atom or a molecule. Here, we refer to these species, which may well be non-native to the tunneling barrier, as "particles under study". An inelastic tunneling process can only occur if $E_i$ exceeds the energy $E_{exc}$ required for the excitation. It is therefore only for $E_i > E_{exc}$ that inelastic tunneling provides a channel for electrons to cross the barrier. Accordingly, the conductance of the tunnel junction increases stepwise at the tunnelling voltage $V = E_{exc}$, and so the second derivative of the junction's current–voltage characteristic $\partial^2 I/\partial V^2(V)$ shows a peak. As this peak is centered at $E_{exc}$, it quantifies the excitation energy and therefore allows the particle under study to be identified. The peak's strength is a measure of the particles' concentration and scattering cross sections.

To explore the potential of IETS to study ionic species at high temperatures, we used the same $BaZrO_3$-based heterostructures to act simultaneously as proton conductors and tunnel barriers. $BaZrO_3$-based multilayers feature excellent dielectric properties (electronic band gap 5.33 eV at 300 K), chemical and thermal stability, and high ionic conductivity.[20,21,22,23,24] Indeed, as proton conductors, such perovskite heterostructures are of great interest for various applications such as solid oxide fuel cells, hydrogen sensors, electrochromic displays, and electrolyzers.[25,26,27,28] Figure 1b shows our fabricated tunnel junctions with Au electrodes grown on top and $SrRuO_3$ on the bottom. In this architecture, the tunnel barriers provide fast diffusion paths for the particles under study—in our case protons or deuterons—into and through the junctions. The species and the quantity of these particles can then be analyzed in real time by IETS with a spatial resolution defined by the size and pitch of the tunnel junctions.

In the experiments, we used Y-doped $BaZrO_3$ films. Doping with trivalent acceptor impurities such as $Y^{3+}$ or $In^{3+}$ on the $Zr^{4+}$ site is the common strategy to create oxygen vacancies that enable dissociate water uptake. The water uptake leads to the formation of internal $OH^-$ groups that give rise to proton conductivities in bulk[29] and thin films[30] exceeding $10^{-2}$ S/cm at 500°C (for 20% Y-doping). In uniformly doped $Ba(Zr_{1-x}Y_x)O_3$ barriers, protons effectively occupy the complete tunnel barrier. This is not always desirable for IETS. To provide a defined local environment for the particles under study and to minimize their interactions with the contacts of the tunnel junctions, we also employed a second approach to enhance proton conduction of $BaZrO_3$ and take advantage of proton diffusion along interfaces in $BaZrO_3$–$BaYO_x$-based multilayers.[31] Strong enhancements of ion conduction at interfaces in heterostructures have already been found for several heterostructures, *e.g.,* for $CaF_2$–$BaF_2$ and $SrCeO_3$–$SrZrO_3$ superlattices.[32,33]

Therefore, the architecture of some of the tunnelling barrier was chosen to take advantage of two-dimensional doping.[31] The tunnel barrier comprises of a $BaYO_x$ layer embedded between $BaZrO_3$ layers (Figure 1b). In such a structure, a $ZrO_2$ layer is replaced in the growth sequence by a $YO_x$ layer, which is intercalated between $BaO_x$ layers, thereby generating oxygen vacancies along the Y-doped layers.[31] In hydrated conditions, hydroxide ions partially fill these oxygen vacancies, yielding the desired proton-conducting channels located in the middle of the barrier. For reference, we also explored tunnel junctions with barriers consisting of undoped and Y-doped $BaZrO_3$ layers.

**Device fabrication and structural characterization**

The devices were fabricated by *in situ* pulsed laser deposition controlled by reflection high-energy electron diffraction and a photolithographic process as described in the supplementary information. To obtain desirable tunnel resistances, the thicknesses of the BaZrO$_3$-based films and heterostructures $d$ were chosen to be between 2.5 and 3 nm, and the tunnel junctions areas $A$ are 10×10, 20×20, and 40×40 µm$^2$. After the photolithography step, which subjected the samples to water and C$_4$H$_{13}$NO-based developer (pH~13), the samples underwent photoresist baking at 125°C. To induce further proton (deuteron) diffusion into the samples, several of the tunnel junctions bonded to their carrier chips were soaked in heated regular water or heavy water at later times (see supplementary information). Figures 1b and 2a show a schematic side view of a device and a top-view scanning electron microscopy image of a tunnel junction, respectively. In total, ten junctions on five substrates were measured: two junctions per substrate for five different substrates labelled A–E.

Characterization of the samples by *x*-ray diffraction and atomic force microscopy reveal a high sample quality (see supplementary information). No indications of pinholes were seen. The cross section of a sample was investigated by scanning transmission electron microscopy (STEM). Figure 2b depicts a high-angle annular dark-field-STEM image of a representative SrTiO$_3$–SrRuO$_3$–BaZrO$_3$–BaYO$_x$–BaZrO$_3$–Au sample. All layers in the stack are coherent despite the large lattice mismatch between SrRuO$_3$ (*a*=3.93 Å) and BaZrO$_3$ (*a*=4.19 Å) which induces compressive strain in the BaZrO$_3$. The barrier thicknesses extracted from STEM data agree well with the *x*-ray reflectivity measurements (4.8 nm vs 4.6 nm for the samples of Figure 2b and Figure S1b, respectively).

**Basic characteristics of the tunnel junctions**

The junctions are highly resistive, and their resistances rise by one order of magnitude if $d$ is enhanced from 2.5 to 3 nm (see supplementary information). The $I(V)$ characteristics of junction C1 ($d$= 2.5 nm, $A$= 20×20 µm$^2$) are shown in Figure S3b for $T$= 2–400 K. These data were taken on the junction as fabricated, *i.e.*, after photolithography but prior to being soaked in H$_2$O or D$_2$O. The maximal voltage range of ±650 mV was chosen to prevent electric breakdown and excessive heating. The curves are nominally symmetric and overlap for $T \leq 150$ K. At higher $T$, the conductance of the junctions increases in accordance with a growing contribution of thermally activated electron transport. These properties match the behavior expected for electron transport by tunneling plus thermally activated electron hopping over the barrier.

### Exploration of inelastic electron tunneling spectra

Having established that the devices act as tunnel junctions, we now present the IETS spectra. IETS measurements were performed between 2 and 400 K. The upper limit was chosen by wiring-induced constraints. The $\partial^2 I/\partial V^2(V)$ characteristics were directly measured using a standard 4-terminal lock-in-amplifier technique that recorded the second harmonic response of the tunnel junctions (see supplementary information). The characteristics were reproducible, also with respect to different deposition runs on different SrTiO$_3$ substrates (**Figure 3**a) and were stable for at least six weeks. Small voltage differences, at which characteristic features occur among different samples, are attributed to the 4-point geometry being not perfect and, possibly, to minor sample heating.

#### 1. Low-temperature IETS spectra

The $\partial^2 I/\partial V^2(V)$ characteristic of sample C1 is shown in **Figure 4**. In forward (positive) bias, electrons tunnel from the bottom SrRuO$_3$ film to the Au top contact. The $\partial^2 I/\partial V^2(V)$ curve shows a rich spectrum of features, which we found reproducibly in all junctions with barriers consisting of Ba(Zr$_{0.8}$Y$_{0.2}$)O$_3$ films or of (BaZrO$_3$)$_2$–(BaYO$_x$)$_1$–(BaZrO$_3$)$_2$ multilayers. The details of these features are also independent of deposition run, SrTiO$_3$ substrate and scan direction (see Figure 3a and supplementary information). However, tunnel junctions with undoped BaZrO$_3$ barriers showed only a minute peak structure (see supplementary information). As the proton conductivity is much lower in BaZrO$_3$ than in Ba(Zr$_{0.8}$Y$_{0.2}$)O$_3$ or BaZrO$_3$–BaYO$_x$–BaZrO$_3$,[20,22,23,24,29,30,31] this observation suggests that the observed peaks are associated with the presence of protons in the barrier.

The spectra reveal three ranges of energy with distinct characteristics: first, a low-energy region (<60 mV) with several small but sharp peaks of $\partial^2 I/\partial V^2(V)$ (Figure 4b), second, an energy window from 60 to 230 mV that is nominally devoid of IETS structures (Figure 4a) and, third, the energy range above 230 mV, which features a rich spectrum of large, crisp peaks (Figure 4c). We now analyze the peaks found in the low and high-energy ranges.

(a) Low-energy range (<60 mV)

As Figure 4b shows, the height of the peaks in this energy range is partly asymmetric with respect to bias direction, whereas the peak energies are symmetric. The asymmetry in the peak height is common for inelastic tunneling processes. The intensity of a peak in an IETS spectrum is high if the inelastic interaction occurs close to the electrode via which the electrons leave the tunneling barrier,[4,34] which for the spectra shown in Figure 4b is the Au contact. The energy of these peaks is in the range of tens of meV, which is the range of phonon energies. Indeed, the two main peaks (15 and 53 mV) labeled 1 and 2 in Figure 4b, match optical phonon modes (14.5 and 55 meV) measured by IR and Raman spectroscopy.[11,35,36] In $BaZrO_3$, the Ba–$ZrO_6$ stretching mode (in our case also the Ba–$YO_6$ stretching) is known to generate a vibrational band at 15 meV, whereas the O–Zr–O bending mode induces a band at 53 meV.[11,36] As expected for phonon-assisted tunneling, the energies of the modes remain unchanged with increasing temperature, and the width of the peaks increases (Figure S6). The peaks start to merge above 10 K.

(b) High-energy range (>230 mV)

In the high-energy range, the IETS spectrum resolves ten pronounced, reproducible peaks (Figure 4c). A comparison with reported IR spectra, INS spectra, and results of *ab initio* molecular dynamics calculations allows us to attribute most peaks to vibrational modes involving hydrogen or to their higher-order transitions.[36,37,38,39] Hydrogen is supposed to be absent in the idealized barrier structure of our samples, but hydration causing proton, hydrogen, hydroxyl, or water diffusion is bound to occur during the photolithography steps and subsequent sample heating. Most peaks observed in Figure 4c match vibrational modes in proton-conducting oxides.[10,36,38] We use the labeling scheme of Figure 4c in this discussion.

The 2δ[O–H] higher-order transitions of the O–H bend mode have been found by INS to occur in In-doped BaZrO$_3$ at 260 and 315 meV. These energies precisely match peaks 3 and 4 of the IETS spectrum.[37] Peak 7 at 450 meV is a peak that has been prominent in the IETS spectra of a variety of metal–oxide–metal tunnel junctions since the 1960s.[1,2,18,40] It has been attributed to a υ[O–H] mode, with free OH⁻ being generated by oxidation processes in the presence of water and subsequent redistribution of hydrogen bonds. As reported in Ref. [37], this 450 meV υ[O–H] mode is usually red-shifted in hydrated oxides. Peak 6 (410 meV) agrees well with this red-shifted mode so that we tentatively attribute it to the υ[O–H] mode arising from hydration in the oxide barrier. Hydration of the junctions is to be expected, as described above. Furthermore, peaks arising from higher-order transitions and combinations of the δ[O–H] and the υ[O–H] modes are anticipated.[37] Indeed, peaks 9–12 at 525, 560, 580, and 635 mV, respectively, exactly match the expected energies of the δ[O–H]+υ[O–H] and 2δ[O–H]+υ[O–H] modes (Figure 4c). We cannot assign the remaining peaks 5 (365 mV) and 8 (510 mV) to any combination of δ[O–H] and υ[O–H] modes reported for acceptor-doped BaZrO$_3$. We therefore suggest that these peaks indicate modes associated with oxygen vacancies or related defects present in the barrier.

It is noteworthy that some of the known O–H bending or stretching modes are weak or even absent from the spectrum. Such a behavior of IETS spectra was already noted and analyzed in the early days of IETS.[41,42] Although no strict selection rules exist for IETS, propensity rules for the peak strengths have been defined that take into account the effects at the position, orientation, and screening of the dipole electric field of the modes interacting with the tunnelling electrons.[41,42,43,44,45,46] In addition, the spectra of our junctions are characterized by the almost featureless energy range from 60 to 230 mV. The peaks expected in this range are muted by an unknown phenomenon.

Isotope replacement is a control experiment capable of revealing the role of protons in vibrational modes. We therefore inserted deuterons into the barriers by subjecting the tunnel junctions to D$_2$O (see supplementary information). Figure 3b shows the $\partial^2 I/\partial V^2 (V)$ characteristics of a SrTiO$_3$–SrRuO$_3$–BaZrO$_3$–BaYO$_x$–BaZrO$_3$–Au tunnel junction (B3, $d$ = 2.5 nm) before and after exposure to deuterium, which induces the marked peak at 328 mV that matches the O–D stretching mode.[47,48] The same peak was reported for deuterated Al–Al$_2$O$_3$–Pb junctions as being the only peak above 230 mV.[2] Its presence confirms that deuterons have penetrated the tunnel barrier and thereby provides evidence that (*i*) protons indeed readily diffuse into the barrier and (*ii*) their vibrational modes generate peaks in the high-energy range of the $\partial^2 I/\partial V^2(V)$ characteristics.

Although one might expect that during the D$_2$O-soaking experiment some of the protons in the OH-groups to be exchanged with deuterons, the IETS spectra reveal no noticeable shifts of the O–H peaks. It seems that the D$_2$O-soaking parameters used did not lead to a general replacement of the protons embedded in the crystal lattices. These results suggest that it is much easier to add protons during sample fabrication than to replace them afterwards.

## 2. IETS spectra as a function of temperature

IETS could be highly useful if it were possible to perform tunneling spectroscopy above room temperature. This requires a spectral resolution sufficient to yield a peak structure in the $\partial^2 I/\partial V^2(V)$ characteristics that allows the modes to be identified and possibly quantified. Figure 4a displays high-resolution IETS spectra of sample C1 taken between 2 and 400 K that clearly show temperature-driven changes in the characteristics. Importantly, however, the spectra show well-resolved sets of peaks at all temperatures, even at 400 K. Many of these peaks are connected to the previously identified low-temperature modes (Figure 4c). The intensity of the phonon peaks (peaks 1 and 2) weakens with rising temperature. Above 100 K, the sharp zero-bias V-shaped depression of the $\partial^2 I/\partial V^2(V)$ characteristic smoothens into a dip. This behavior is expected from the thermal softening of the electrodes' Fermi–Dirac distributions.[8,16]

Above 200 meV, the measured IETS peaks, which reflect the O–H vibration bands, persist up to 400 K (Figure 4a). Starting at 150 K, the spectra show that the O–H modes shift by ~10 mV with temperature increasing to 400 K (for peak 3, *e.g.*, by 15 mV). The mechanism causing the observed peak shift, which at its face value would imply a thermal softening of the O–H stretching modes, remains to be investigated.

The peak widths of the 450 meV υ[O–H] modes (15.2 meV) at low temperatures are much smaller than the corresponding peak widths obtained by IR and Raman measurements on polycrystalline $BaZrO_3$-based bulk samples, but are comparable to the IETS peak widths obtained at Al–$Al_2O_3$–Pb junctions (~17 meV).[2,48] It is reasonable to expect that the peak widths of our samples are smaller than those of the bulk polycrystals because the epitaxially oriented, few-unit-cell thick tunnel barriers provide a more regular microstructure.

We did not expect that the IETS spectra would exhibit spectral peaks that are sufficiently distinct at 400 K to identify the presence of protons at interfaces. Indeed, this finding disagrees with the Lambe–Jaklevic limit of the temperature-dependent spectral resolution of 5.4 $k_BT$ for the minimal peak width (FWHM). As Figure 3c shows, the measured peak widths, determined as described in the supplementary information, indeed violate this limit above a sample-specific temperature of 30–150 K. This is shown most pronouncedly by sample C1. For this sample, the width of the 580 meV mode (peak 11) equals 20.1 meV at 400 K, which is a factor of 9 smaller than the 5.4 $k_BT$ limit of ~186 meV. At 300 K, the peak width equals ~19 meV, corresponding to 0.75 $k_BT$. Such peak widths were found in most—but not all—of the samples investigated. For example, the peak widths of sample D1 are larger (75.5 and 81 meV at 300 and 400 K, respectively, see Figure 3c), but still less than half of the Lambe–Jaklevic limit. We attribute the observed resolution enhancement to a spectral sharpening of the tunneling electrons induced by sample-dependent resonant states. In this hypothetical scenario, the standard energy distribution of the tunneling electrons is sharpened by resonant electron tunneling, which occurs in addition to direct tunneling, in case the resonant states have a spectral width that is much smaller than the thermal energy distribution of the electrons in the electrodes (see Figure S7 in supplementary information). This resonant tunneling is reminiscent to the previously reported orbital-mediated tunneling through molecules.[49,50] In our case, defects of the oxide-crystal lattices or interface states may provide for the resonant states. With this high resolution, IETS has the capability to yield valuable spectral information even at temperatures above 400 K. To achieve such high operation temperatures requires further optimization of the junctions' thermal stability, including their contacts and wiring.

Having demonstrated that IETS with planar tunnel junctions is a viable high-resolution technique to analyze mobile species in solids, we draw attention to the fact that this technique is only useful to perform spectroscopy on solids that can be integrated in tunneling barriers. This implies that the materials must be excellent insulators, even at layer thicknesses as small as several nanometers. There is no explicit need for the layers to be grown with epitaxial order. The requirement to embed the materials of interest in tunnel junctions implies that IETS is not a suitable technique to study test samples of random bulk materials. Nevertheless, we foresee IETS as a highly valuable tool to analyze materials and heterostructures grown as thin layers specifically for the task of exploring their ionic dynamics. In this case, lithography with all its available freedom for design can be used to design experimental setups. **Figure 5** shows a corresponding, envisaged nanoionic device that utilizes nanostructured tunnel junctions for time resolved, 2D mapping of ionic diffusion in heterostructure devices.

In conclusion, we have designed and validated tunnel junctions in which the tunnel barrier consists of efficient proton conductors. The barriers of these junctions therefore have a double function. They act as a vertical tunneling barrier for electrons and as a horizontal diffusion path for ions. The tunnel junctions yield meaningful IETS spectra of diffused ions up to at least 400 K with no upper bound identified. The excellent fit of the peak energies in the IETS spectra enables them to be attributed to established phonon modes and to O–H bend and stretch modes, which has been confirmed by spectroscopy of tunnel junctions subjected to $D_2O$.

This work provides evidence that, despite a well-accepted tenet to the contrary, IETS is a powerful spectroscopic tool, capable to analyse ions above room temperature. At 400 K, the achieved spectral resolution of the junctions is at least a factor of 9 better than the established theoretical limit. Applying IETS to ionic conductors opens the door to studying diffusion in solids in real time with a spatial resolution limited only by the minimum size of the tunnel junctions. We expect future IETS to utilize even smaller tunnel junctions, including junctions fabricated from 2D van-der-Waals materials, to operate at even higher temperatures, and to monitor the motion of several ionic species, such as protons, lithium, and sodium ions simultaneously. With the capability of high-temperature operation, IETS has the potential to become a valuable item in the suite of analytical tools for future solid-state ionics.

**Supporting Information**

Supporting Information is available from the Wiley Online Library or from the authors.

**Acknowledgements**

We acknowledge valuable discussions in particular with R. Merkle, H. Boschker, D. Braak, P. Bredol, S. Kivelson, T. Harada, and T. Whittles. We thank M. Hagel and H. Hoier for technical support, and L.-M. Pavka for editorial help. Y. Wang and P. A. van Aken acknowledge funding from the European Union's Horizon 2020 research and innovation programme under grant agreement No. 823717 – ESTEEM3.

**Conflict of interest:** The authors declare no conflict of interest.

Received: ((will be filled in by the editorial staff))
Revised: ((will be filled in by the editorial staff))
Published online: ((will be filled in by the editorial staff))

*Figures and Captions:*

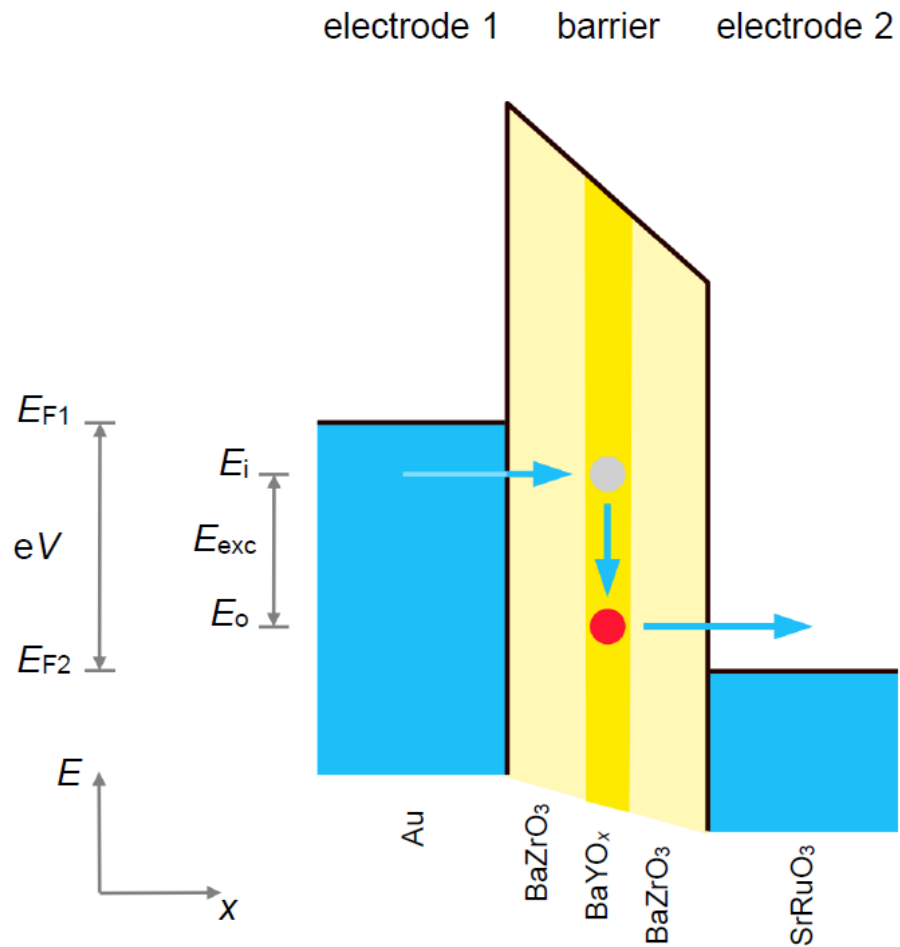

**Figure 1**

(a) Schematic illustration of an inelastic electron tunnelling process. Owing to an applied current, an electron leaves electrode 1 and enters the tunnel barrier with an energy of $E_i$. In the barrier, the electron transfers $E_{exc}$ of its energy to the particle of interest. The electron reaches electrode 2 with an energy of $E_o = E_i - E_{exc}$. The excitation energy $E_{exc}$ is a characteristic property of the particle of interest, and therefore allows it to be identified and characterized. In this example, the tunneling barrier consists of a $BaZrO_3$–$BaYO_x$–$BaZrO_3$ trilayer. The embedded $BaYO_x$ layer is a diffusion channel for protons. The energy spectrum of the vibrational modes caused by the protons present in the $BaYO_x$ layer can be derived by measuring $E_{exc}$.

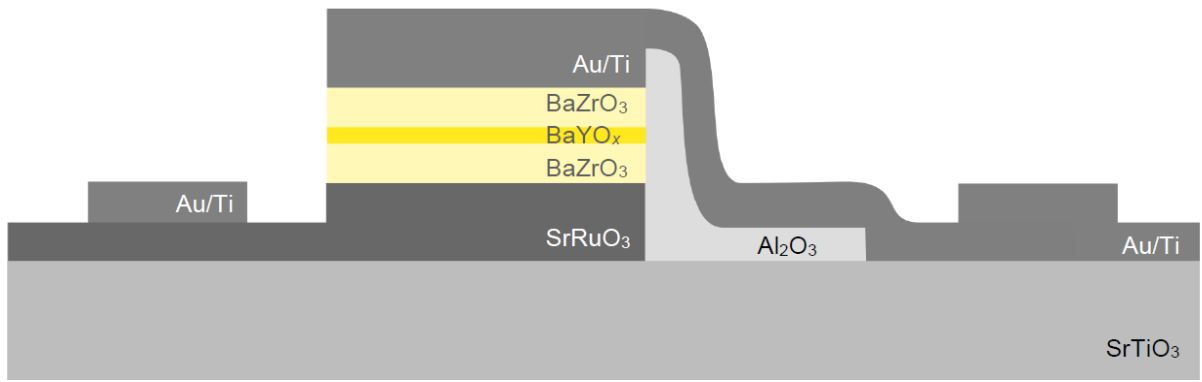

**Figure 1**

(b)  Cross section of a tunnel junction according to (a) (not to scale).

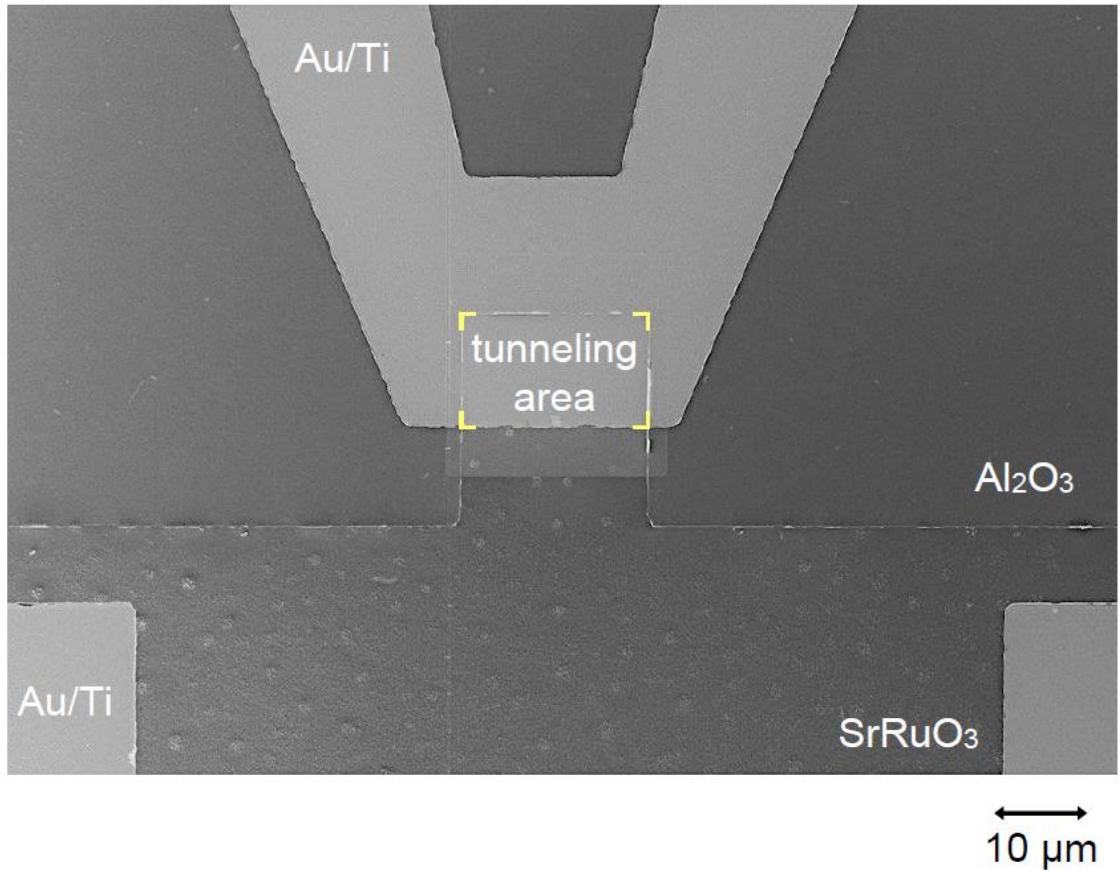

**Figure 2**

(a) Top view scanning electron microscopy image of a typical tunnel junction. The device shown has a proton-conducting $BaZrO_3$–$BaYO_x$–$BaZrO_3$ barrier with an area of 20×20 µm$^2$. The yellow marks indicate the corners of the tunnel area.

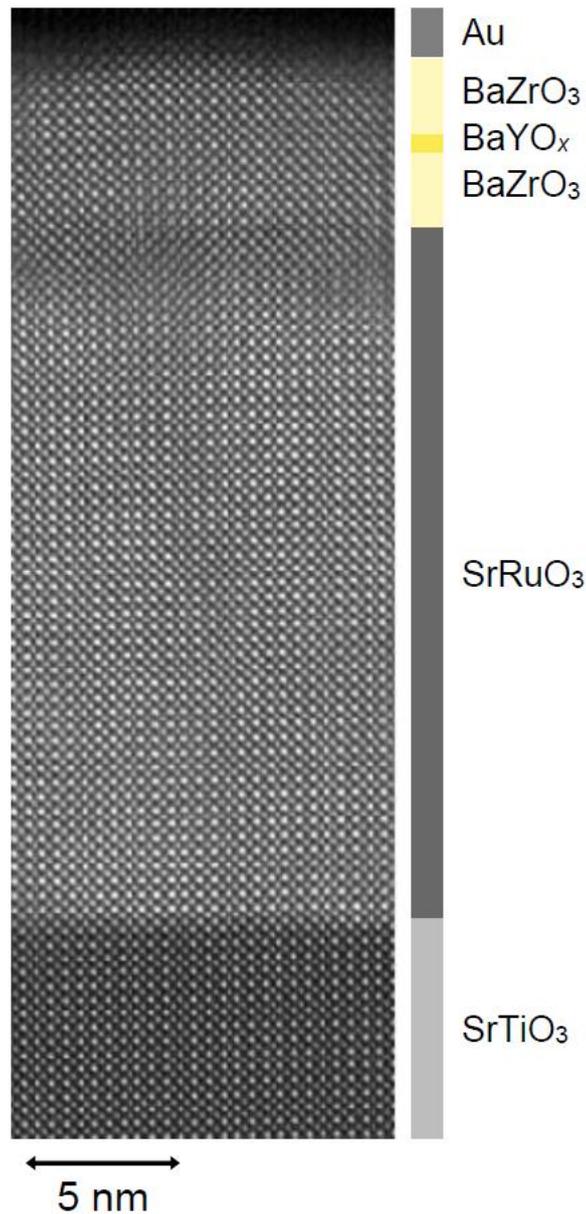

**Figure 2**

(b) High-angle annular dark-field STEM image showing a cross-sectional cut through a $SrTiO_3$–$SrRuO_3$–$BaZrO_3$–$BaYO_x$–$BaZrO_3$–Au tunnel junction. Epitaxial growth has been maintained through the entire stack, and the thicknesses of the individual layers match the intended device structure. The darkish band in the upper part of the $SrRuO_3$ layer is attributed to local strain generated by misfit dislocations generated at the interface, as electron energy loss spectroscopy (EELS) O-K edge intensity distribution maps did not show a dip of the oxygen concentration (see Figure S2 in supplementary information).

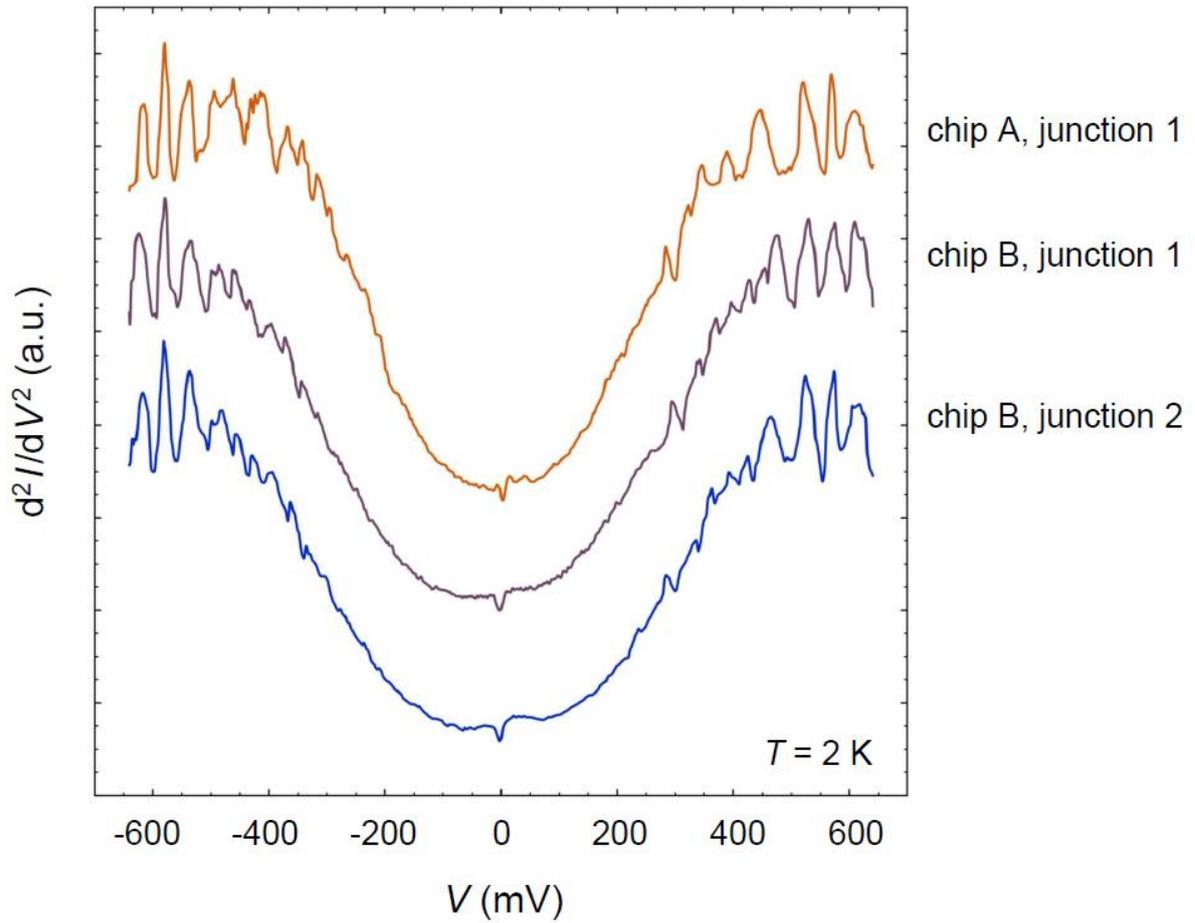

**Figure 3**

(a) IETS spectra of three 2.5-nm-thick $SrTiO_3$–$SrRuO_3$–$BaZrO_3$–$BaYO_x$–$BaZrO_3$–Au tunnel junctions measured at 2 K. The positive bias corresponds to electrons tunneling into the Au top contact. Chips A and B were subject to different deposition runs. For clarity, the curves have been shifted along the *y*-axis.

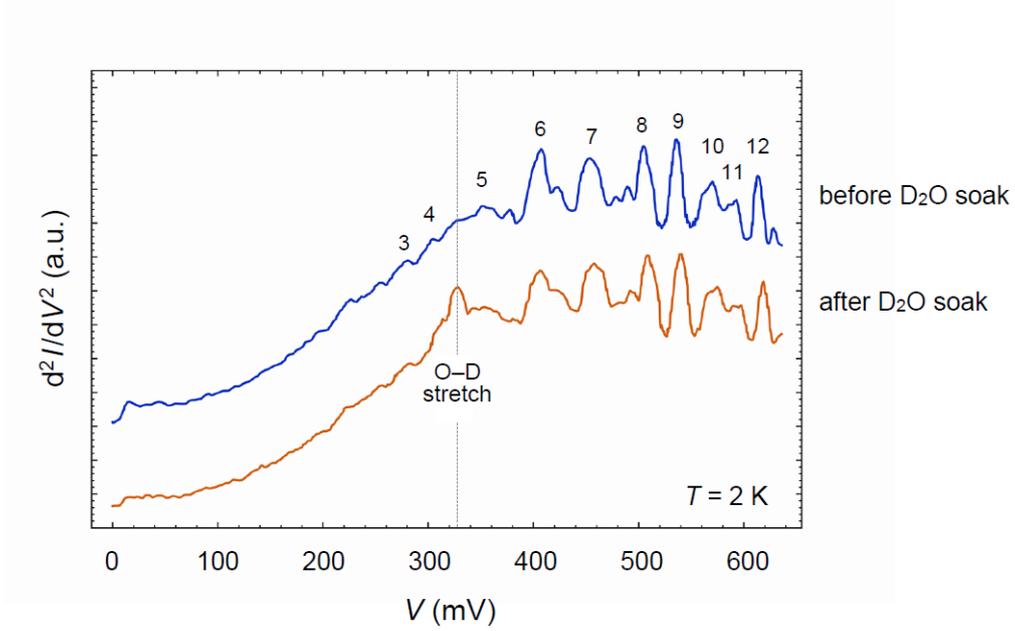

**Figure 3**

(b) Comparison of the IETS spectra of a $SrTiO_3$–$SrRuO_3$–$BaZrO_3$–$BaYO_x$–$BaZrO_3$–Au tunnel junction (sample B3) before (blue) and after (orange) being soaked in $D_2O$. The 328 meV peak generated by the $D_2O$ soak reflects the O–D stretching mode. For clarity, the curves have been shifted along the *y*-axis.

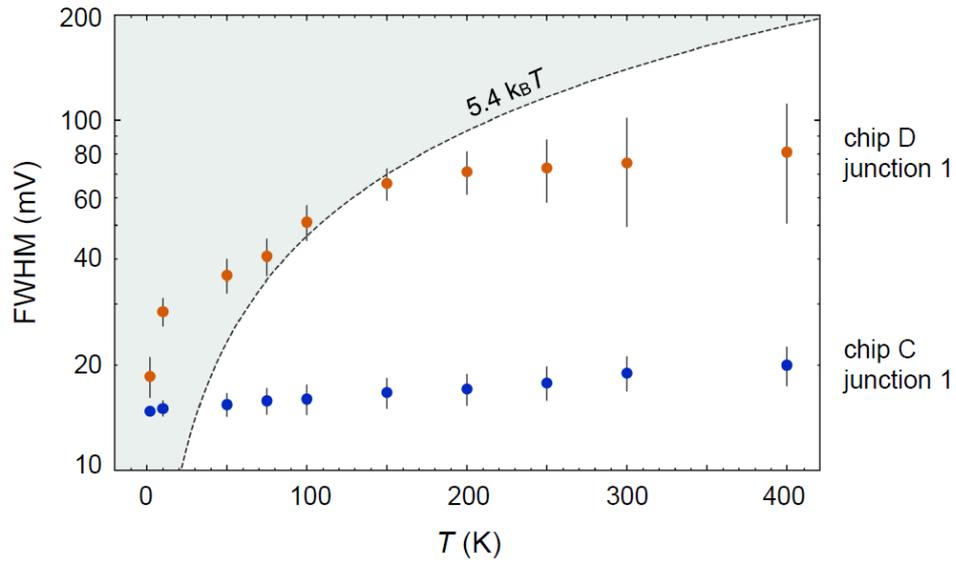

**Figure 3**

(c) Full-width-at-half-maximum peak width of the 580 meV mode (peak 11 in Fig. 4c) for two samples C1 and D1 as a function of temperature. Junction D1 is the junction with the broadest linewidth of all. Peak widths outside the green area are below the Lambe–Jaklevic limit of 5.4 $k_B T$ (hatched line).

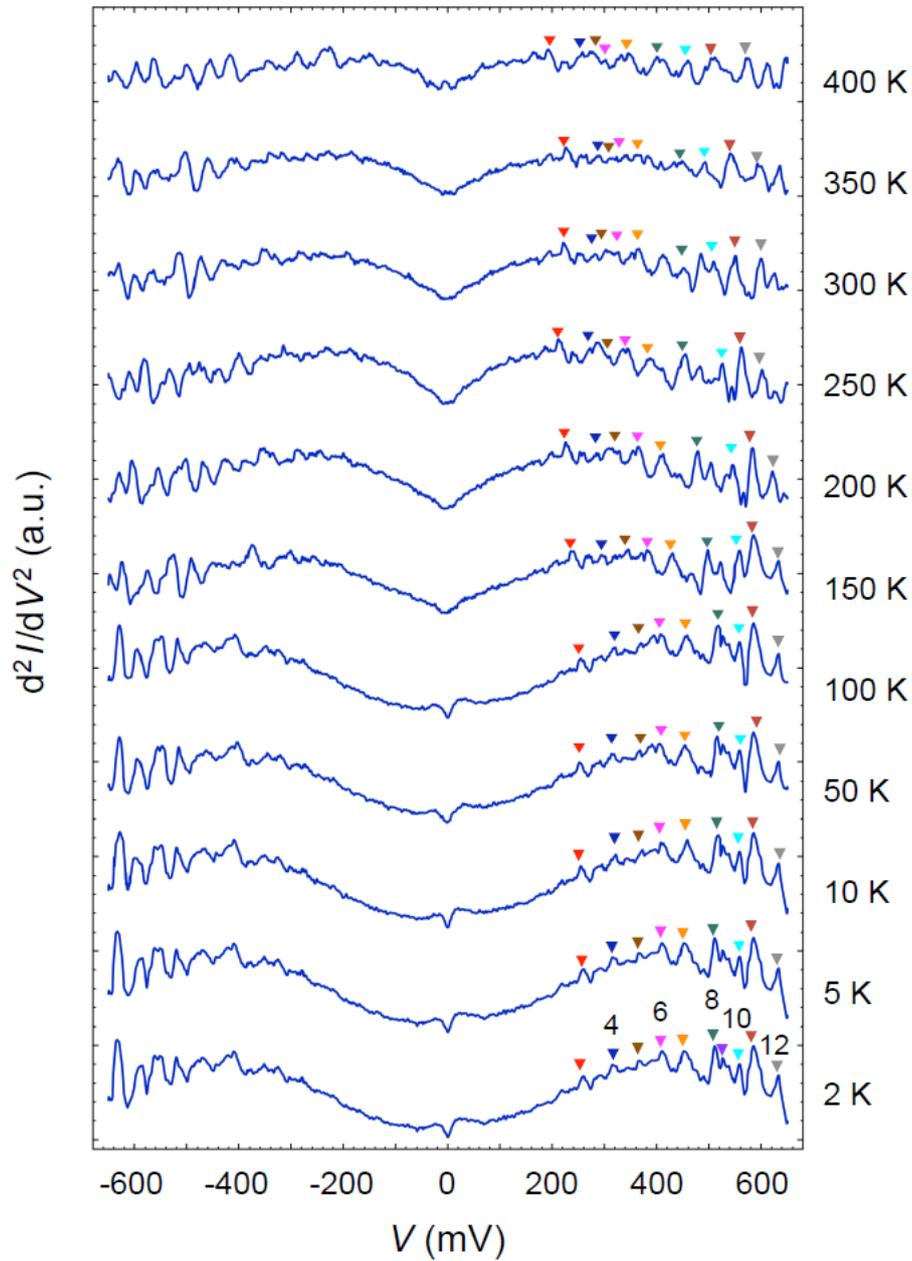

**Figure 4**

(a) IETS spectra of sample C1 measured at a range of temperatures. The low-energy IETS features (<60 meV, peaks 1 and 2) weaken and soften with increasing temperature, giving way to a rounded dip. The high-energy features (>230 meV) are well resolved and traced up to 400 K, the highest temperature used. The curves have been shifted along the *y*-axis for clarity.

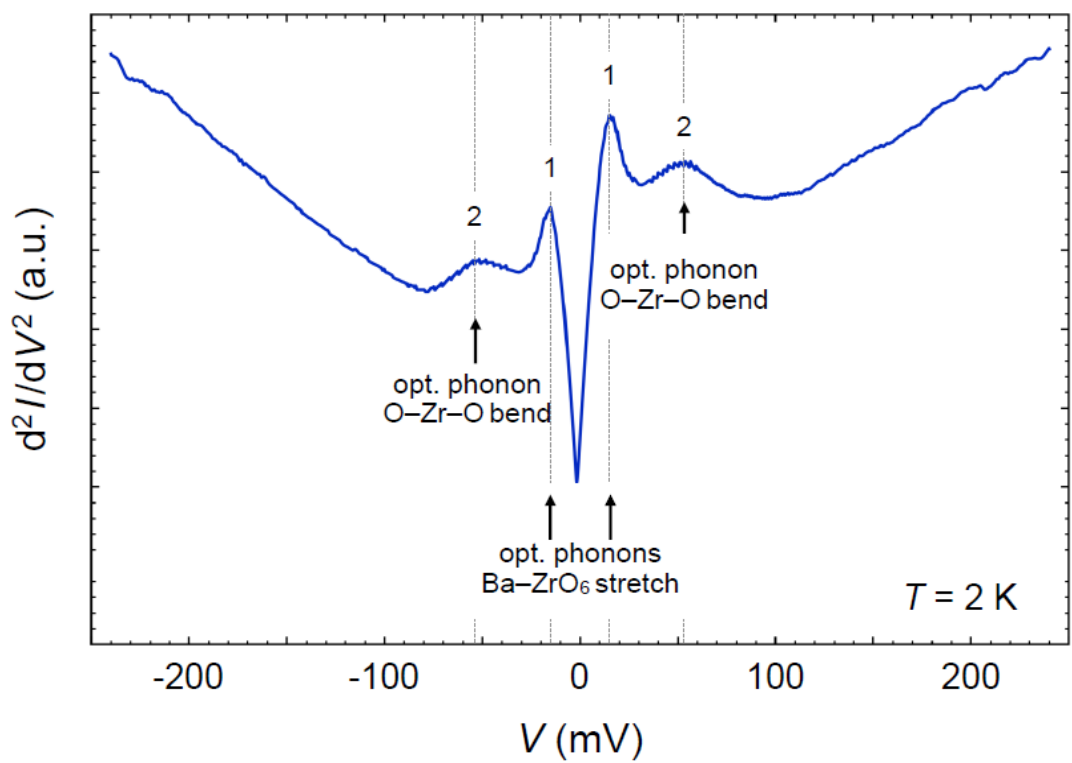

**Figure 4**

(b) Low-energy IETS spectrum of sample C1 measured at 2 K revealing peaks 1 and 2 (15 and 53 meV) resulting from two optical phonon modes in $BaZrO_3$. The V-shape dip at $V \approx 0$ has only been observed in samples containing Y (see supplementary information).

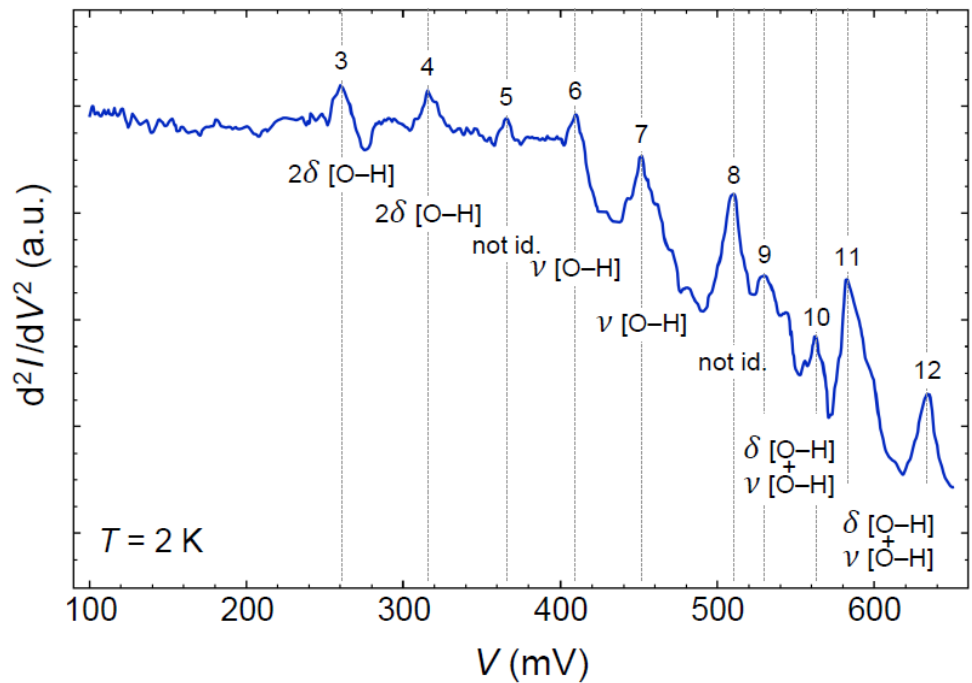

**Figure 4**

(c) High-energy IETS spectrum of sample C1 measured at 2 K displaying the assigned vibrational modes. Peaks 5 and 8 have not been identified.

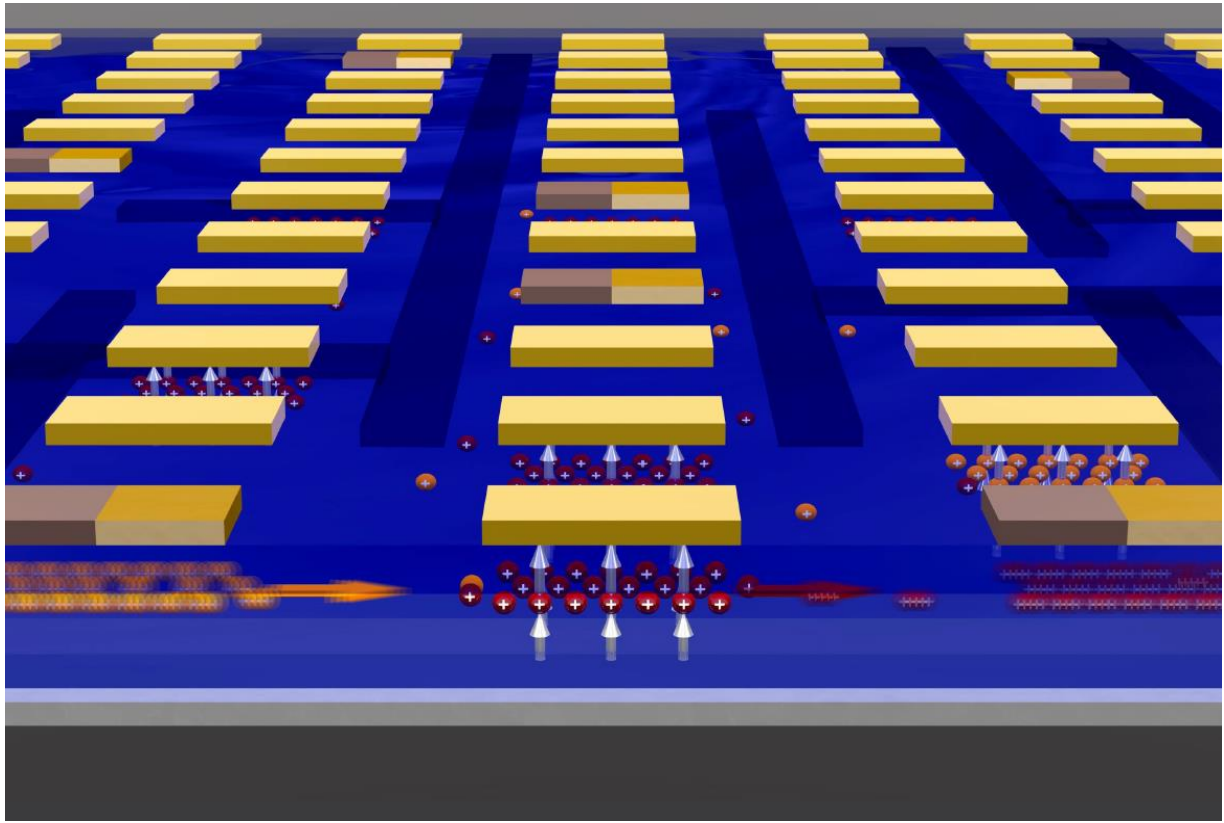

**Figure 5**

    Illustration of a nanoionic device with tunnel sensors monitoring the diffusion of two species of ions (orange and red spheres) along ion-conducting passages (bright blue). Ions diffusing in plane are detected by transversely tunneling electrons (silverish arrows). The ions induce inelastic electron tunneling, which causes ion-specific features in the electron tunneling characteristic. These can be monitored and spatially resolved in real time. The bicolored brown bars depict electrodes. Electric voltages applied between their two halves and voltages applied between two halves of a bottom electrode (not shown) pump the ions in plane.

# Supporting Information

## Inelastic Electron Tunneling Spectroscopy at High-Temperatures

*Prosper Ngabonziza[1], Yi Wang[1], Peter A. van Aken[1],*

*Joachim Maier[1], and Jochen Mannhart[1,]**

[1]Max Planck Institute for Solid State Research,

Heisenbergstraße 1, 70569 Stuttgart, Germany

### Sample preparation and characterization

Samples were grown by pulsed laser deposition ($\lambda= 248$ nm) at a target-substrate distance of 56 mm using a $CO_2$ laser substrate heating system. Prior to deposition, the (100) oriented $SrTiO_3$ substrates were terminated *in situ* at 1300°C for 200 seconds using a $CO_2$ laser.[1] For growth of the tunnel junction stacks, first an epitaxial $SrRuO_3$ layer of a thickness of 60 nm was grown on $SrTiO_3$ using a KrF excimer laser. We used a laser fluence of 2.3 J/cm$^2$ and $8.5\times10^{-2}$ mbar of $O_2$. Subsequently, epitaxial ultrathin layers of $BaZrO_3$–$BaYO_x$–$BaZrO_3$ were grown on $SrRuO_3$ by ablating alternately the $BaZrO_3$ and $BaYO_x$ targets to embed a $BaYO_x$ layer in-between $BaZrO_3$ layers. i.e. $(BaZrO_3)_n$–$(BaYO_x)_1$–$(BaZrO_3)_n$. The barrier layers were grown at $1.3\times10^{-2}$ mbar of $O_2$ using a laser fluence of 2 J/cm$^2$. $SrRuO_3$ bottom electrode and tunnel barriers were deposited at a substrate temperature of 700°C with a laser repetition rate of 1 Hz. Immediately after growth, samples were cooled down to room temperature at 2 K/s for the subsequent *in situ* growth of the Au top electrode. This Au top layer (15 nm) was deposited *in situ* at room temperature, in the same PLD chamber to avoid surface contamination, using a laser fluence of 2.7 J/cm$^2$ at a repetition rate of 30 Hz, in 0.1 mbar of Ar pressure.

---

* Corresponding author: office-mannhart@fkf.mpg.de

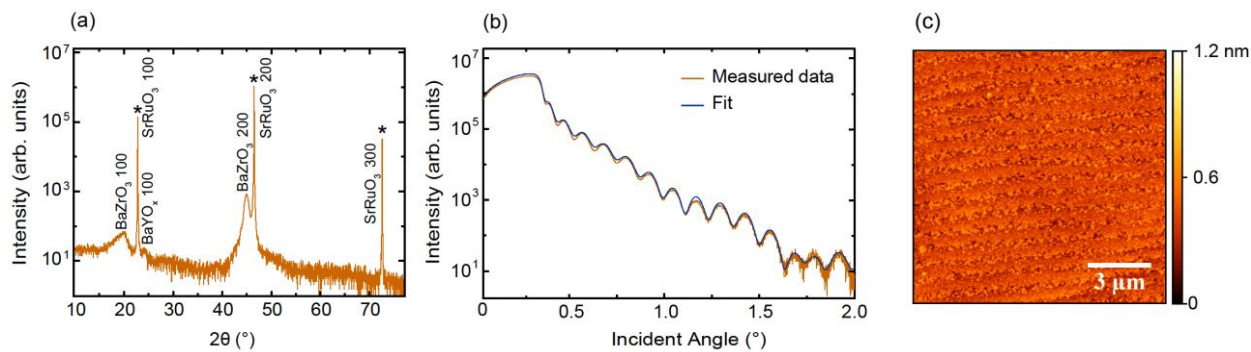

**Figure S1.** (a) *X*-ray diffraction pattern (2θ-ω scan) of a representative SrRuO$_3$–BaZrO$_3$–BaYO$_x$–BaZrO$_3$ heterostructure grown on (100) oriented SrTiO$_3$ substrate. Only the substrate peaks (*) and the (00*l*) family diffraction peaks of all epixatial layers are resolved. (b) *X*-ray reflectivity data of the same heterostructure exhibiting distinct interference fringes. (c) Atomic force microscopy image of the same heterostructure. The extracted root-mean-square surface roughness is ≤0.3 nm over a 10×10 µm$^2$ area.

The scanning transmission electron microscopy (STEM) investigations were performed using a Cs-probe-corrected JEOL JEM-ARM200F.

From the *x*-ray diffraction pattern of a representative SrTiO$_3$–SrRuO$_3$–BaZrO$_3$–BaYO$_x$–BaZrO$_3$ sample (**Figure S1**a), only the substrate peaks and the (00*l*) family of diffraction peaks are resolved; indicating an excellent epitaxial orientation. The thicknesses of individual layers are obtained by fitting simulated reflectivity curves to the measured *x*-ray reflectivity data (Figure S1b). Kiessig fringes well-defined over more than five orders of magnitude in intensity are observed by *x*-ray reflectivity, evidencing homogeneous heterolayers with a smooth surface. Electron energy loss spectroscopy mapping showed that the Y of YO$_2$ monolayers to be confined within a three unit-cell-thick block (see **Figure S2**). This morphology is also confirmed by visibly smooth terraces observed by atomic force microscopy (Figure S1c); thus, indicating the formation of insulating barriers without pinholes.

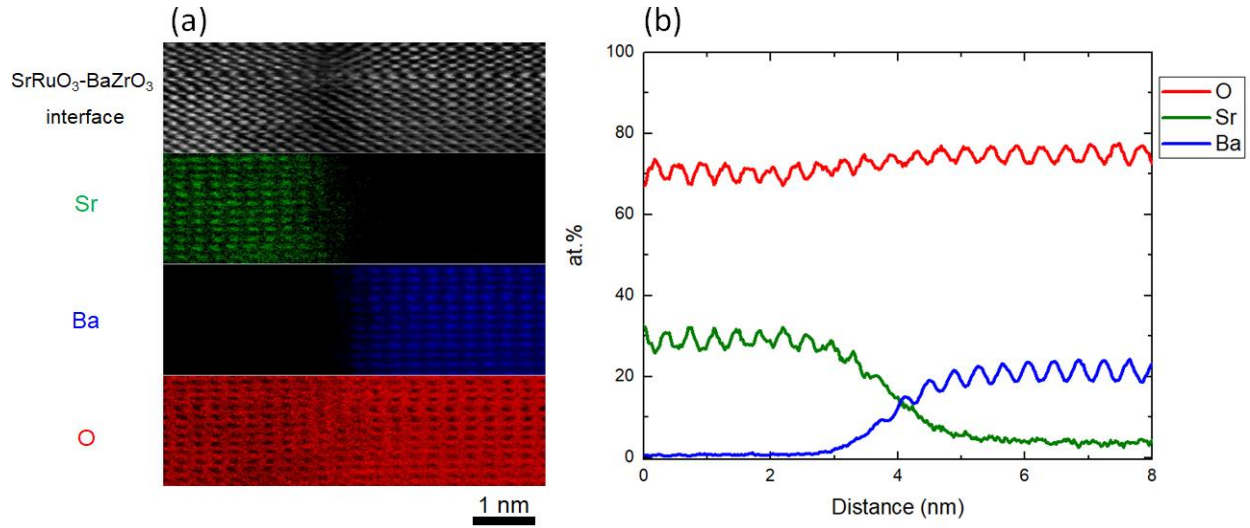

**Figure S2**. (a) Representative cross-sectional high-angle annular dark-field (HAADF) image of the $SrRuO_3$-$BaZrO_3$ interface simultaneously recorded with electron energy-loss spectroscopy (EELS) mapping. Atomic column resolved Sr-L edge (green), Ba-M edge (blue), and O-K edge (red) elemental maps obtained by fitting the EELS data to the reference spectra using a multiple linear least-squares fitting procedure. (b) Averaged line profiles from the EELS spectrum imaging, O-K edge profile shows there is no dip of the oxygen concentration at the interface.

**Basic electrical characteristics of the tunnel junctions**

After growth, the $SrRuO_3$–$BaZrO_3$–$BaYO_x$–$BaZrO_3$–Au structures were patterned into junctions using standard photolithography methods and subsequent ion-beam etching. To separate the top Au electrode from the $SrRuO_3$ bottom electrode, we used a 40 nm $AlO_x$ layer grown by atomic layer deposition. Additional Au/Ti contacts (45 nm thick) that provide electrical connections to the sample were added in subsequent lithography and electron beam evaporation steps.

Basic electrical characterizations of the tunnel junctions were performed after photolithography but prior to being soaked in $H_2O$ or $D_2O$. Samples were measured using a physical property measurement system (PPMS) in four-point configuration utilizing both the PPMS built-in electronics and a source measurement unit (Keithley 6430).

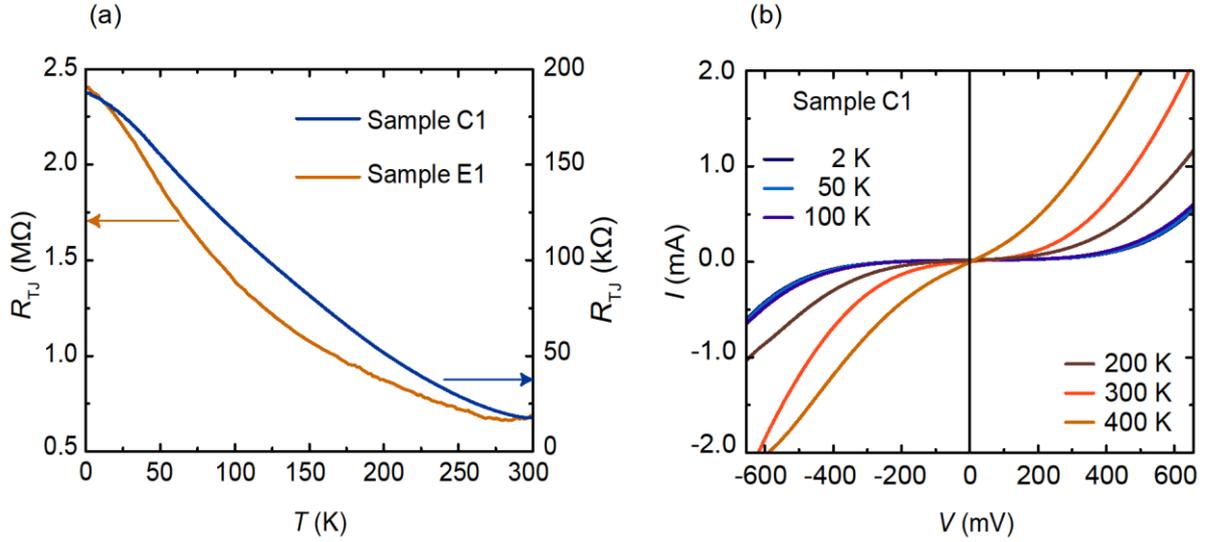

**Figure S3.** (a) Temperature dependence of the junction resistance for two SrRuO$_3$–BaZrO$_3$–BaYO$_x$–BaZrO$_3$–Au samples of the same junction area ($A$ = 20×20 µm$^2$) with different barrier thicknesses: 2.5 nm (sample C1) and 3 nm (sample E1). The samples were measured using an excitation current of 0.5 µA. (b) $I(V)$ characteristics of the sample C1 measured in dc mode at various temperatures.

The junction resistance of two SrRuO$_3$–BaZrO$_3$–BaYO$_x$–BaZrO$_3$–Au devices with barrier thickness of 2.5 nm (sample C1) and 3 nm (sample E1) are shown in **Figure S3**a. The junctions are highly resistive. Figure S3b depicts the $I(V)$ characteristics of the sample C1 at various temperatures.

### Inelastic electron tunnel spectroscopy measurements

For the inelastic electron tunnel spectroscopy (IETS) measurements, we used a lock-in amplifier and a high-frequency current amplifier to apply a small sinusoidal signal, modulate the voltage across the device and to analyze the current response through the tunnel device.[2,3] A low-noise preamplifier was used to monitor the dc voltage across tunnel junctions. IETS spectra were taken with an ac amplitude of 1 mV at a frequency of 1 kHz. $I(V)$ characteristics measured in dc mode compared well to those integrated from the 2$^{nd}$ harmonic.

**Figure S4** illustrates the IETS process, with the sharp increase of the tunneling conductance occurring at $V = E_{exc}$.

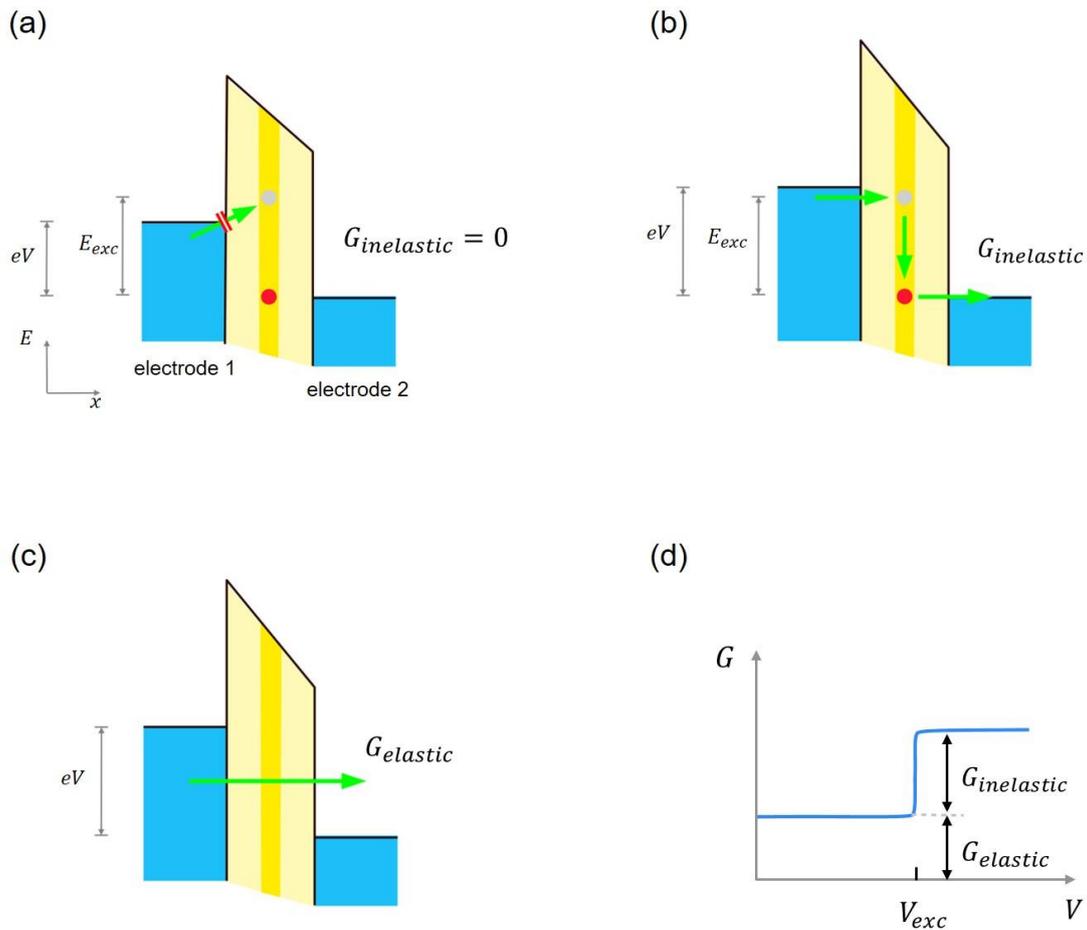

**Figure S4.** (a, b) Band diagram of a tunnel junction at *T*=0 with an inelastic state in the barrier that has a transition energy $E_{exc}$. In panel (a) the junction is biased with a voltage $V < E_{exc}/e$. In this case, electrons cannot tunnel via the inelastic state, because the electrons lack the energy to reach the excited state and after relaxation get to an unoccupied state in electrode 2. If $V$ is enhanced beyond $V_{exc} = E_{exc}/e$, tunneling of electrons via this inelastic transfer becomes possible (b). The conductance of the tunnel junction therefore increases stepwise if $V$ is enhanced beyond $V_{exc}$. Below this voltage, the conductance is due to elastic tunneling only, $G_{elastic}$ (c); above $V_{exc}$ it is the sum of the elastic ($G_{elastic}$) and of the inelastic ($G_{inelastic}$) conductance processes (d).

To induce proton (deuteron) diffusion into the samples, the tunnel junctions bonded to their carrier chips were soaked in regular water or heavy water (open 150 ml beakers, 20 °C, 1 day, H₂O: deionized, D₂O: 99.8 at % D), then heated in the bath for 1 hour at 80 °C. For sample treated in regular water, we have not resolved any noticeable changes in the IETS spectra after $H_2O$ treatment; whereas there is appearance of the O-D stretching mode at 328 meV in the IETS spectra after treatment in $D_2O$ (see Figure 3b in the main text).

To explore the doping effect further in IETS spectra, we have investigated the effects of yttrium doping of the tunnel barriers and the associated proton incorporation on the inelastic tunnel spectra of $Ba(Zr_{0.8}Y_{0.2})O_3$ and undoped $BaZrO_3$ barriers between 2 K and 400 K. For barriers of the same thicknesses ($d=$ 2.5 nm) and comparable resistances (**Figure S5**a), we resolve clear enhanced IETS peaks in the O-H band region only for the $BaZrO_3$–$BaYO_x$–$BaZrO_3$ and $Ba(Zr_{0.8}Y_{0.2})O_3$ barriers (Figure S5b-d).

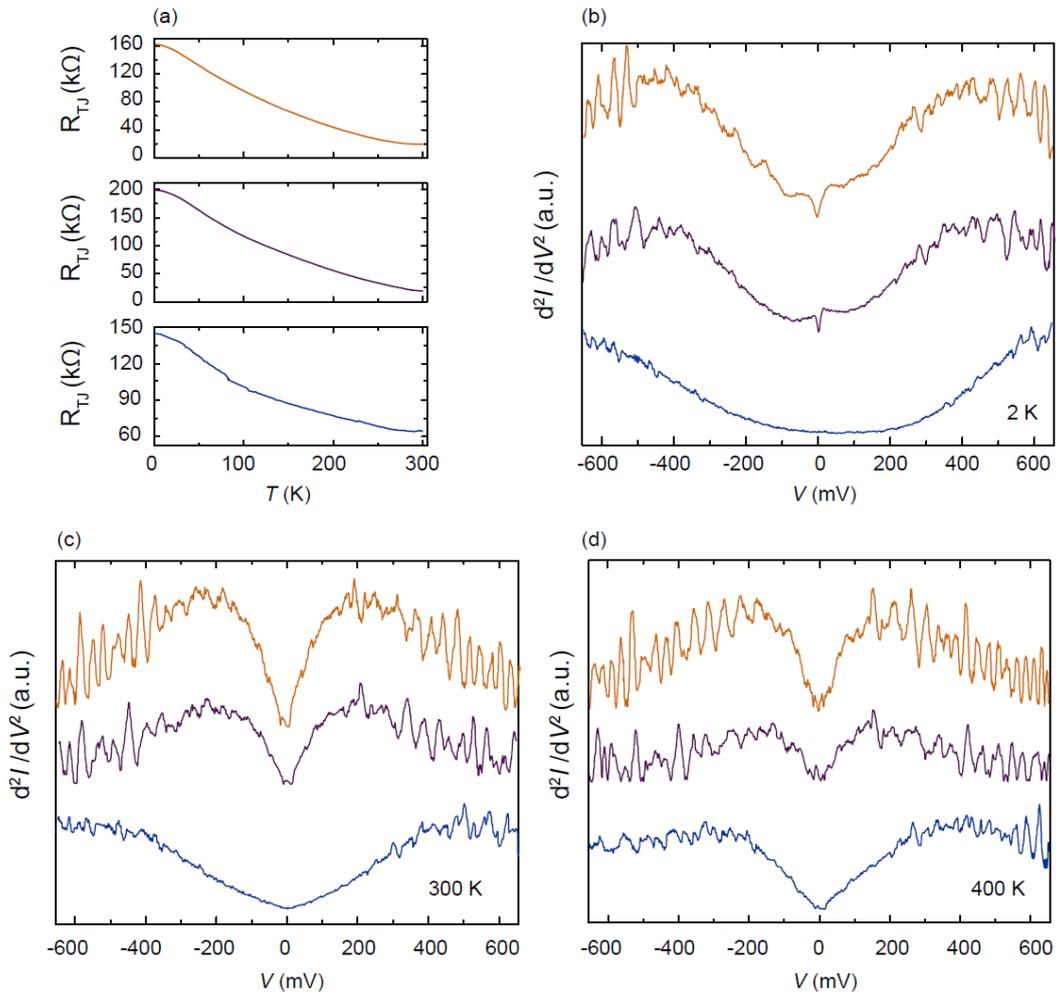

**Figure S5.** Doping effect in IETS spectra for three different tunnel junctions: $SrRuO_3$–$BaZr_{0.8}Y_{0.2}O_3$–Au (orange), $SrRuO_3$–$BaZrO_3$–$BaYO_x$–$BaZrO_3$–Au (dark purple) and $SrRuO_3$–$BaZrO_3$–Au (blue). The thickness of the insulating tunnel barrier is 2.5 nm for all the three junctions. (a) Temperature dependence of the junctions' resistance and (b)-(d) IETS spectra measured at several temperatures.

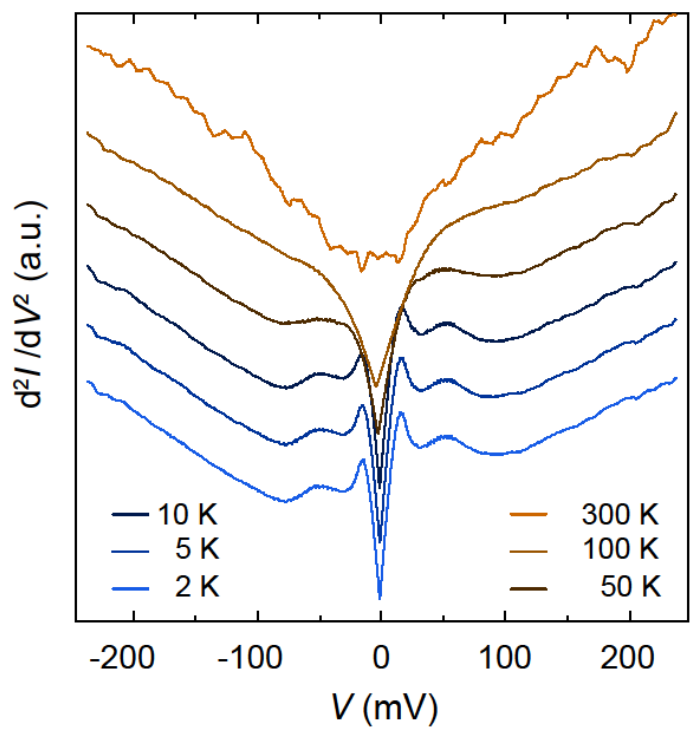

**Figure S6**. High resolution IETS spectra in the low voltage regime of the sample C1 measured at a range of temperatures.

**Figure S6** depicts the temperature dependence of the IETS spectra for the low energy peaks (15 and 53 meV). The widths of these peaks broaden with increasing temperature, which is consistent with phonon-assisted tunneling. To extract the peak widths, full-width-at-half-maximum and the corresponding error range, the linearly extrapolated background was subtracted from the experimental line shape of the peak of interest and the resulting spectrum then fitted with a Gaussian distribution function.[4,5]

**Figure S7** illustrates a mechanism that we propose as a candidate to enhance the energy resolution of IETS to be better than 5.4 $k_B T$. The standard energy distribution of the tunneling electrons is sharpened by resonant electron tunneling, which occurs in addition to direct tunneling, in case the resonant states have a spectral width that is smaller than the thermal energy distribution of the electrons in the electrodes. This resonant tunneling bears a close similarity to orbital-mediated tunneling through orbitals of molecules embedded in tunnel junctions.[6,7] The resonant states likely depend on sample growth parameters, which matches the observation that the sharpening is sample dependent.

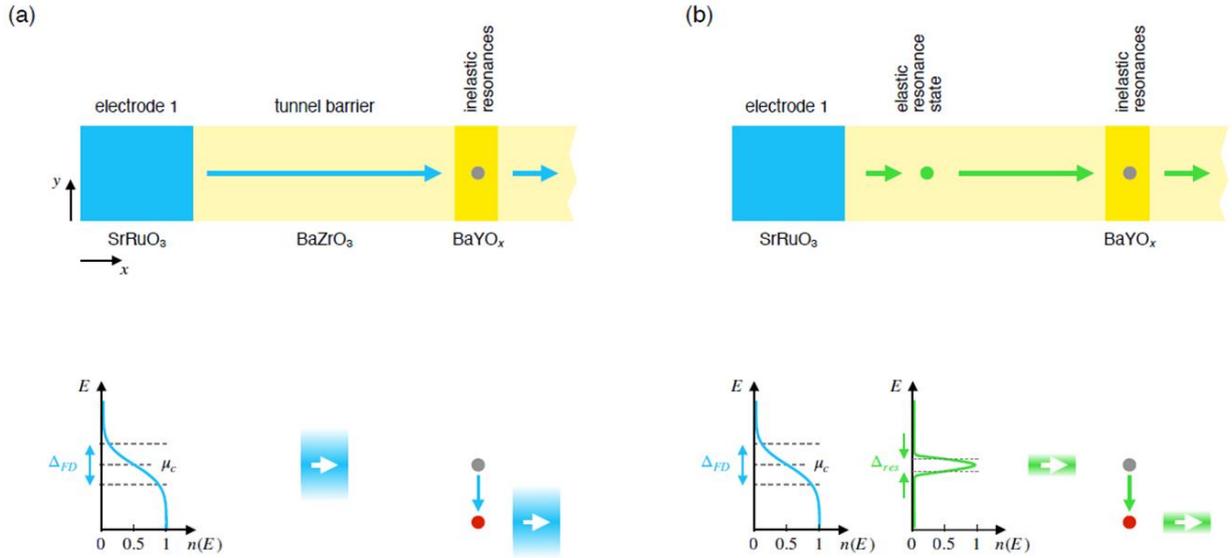

**Figure S7.** Illustration highlighting the difference between conventional inelastic tunneling (a) and the tentatively proposed tunneling mechanism causing an enhanced spectral resolution in inelastic tunneling spectroscopy (b). Both (a) and (b) show on top the implementation of such tunneling barriers using the BaZrO$_3$-based junctions discussed in this work, and at the bottom the corresponding spectral distribution of the tunneling electrons. In (a), the thermal energy distribution (Fermi-Dirac distribution) of the electrons in electrode 1 generates a correspondingly broad energy distribution of the tunneling electrons $\Delta_{FD}$, which, together with the energy distribution of electrode 2, yields an effective spectral resolution of inelastic resonances in the barrier of 5.4 k$_B$T. In the proposed mechanism shown in (b) elastic resonant states with a spectral width $\Delta_{res} \ll \Delta_{FD}$ are embedded in the crystal lattice close to the electrode(s) or at the interfaces between the electrode(s) and the tunneling barrier. The energy distribution of the electrons tunneling via these resonant states corresponds to $\Delta_{res}$ and therefore yields a spectral resolution better than 5.4 k$_B$T.

The suggested mechanism agrees with the fact that the sharpening sets in at high *T*. Furthermore, we note that complex electrode materials with nontrivial band structures may feature at the chemical potential a peak in the density of states oriented towards the tunneling barrier, which may cause an effective sharpening of the tunneling electrons' spectral distribution.

**Table S1.** Full-width-at-half maximum resolution of peak 11 (Figure 4c in the main text) measured at 400 K for all samples.

| Sample | FWHM at 400 K |
|:---:|:---:|
| A1 | 21.2 meV |
| A2 | 21.5 meV |
| B1 | 20.5 meV |
| B2 | 20.9 meV |
| C1 | 20.1 meV |
| C2 | 20.5 meV |
| D1 | 81 meV |
| D2 | 80.5 meV |
| E1 | 43.2 meV |
| E2 | 43.0 meV |